# Electrical switching of two-dimensional van der Waals magnets

Shengwei Jiang[1,2], Jie Shan[1,2,3]*, and Kin Fai Mak[1,2,3]*

[1]Department of Physics and School of Applied and Engineering Physics, Cornell University, Ithaca, New York 14853, USA
[2]Department of Physics, The Pennsylvania State University, University Park, Pennsylvania 16802, USA
[3]Kavli Institute at Cornell for Nanoscale Science, Ithaca, New York 14853, USA

*Correspondence to: jie.shan@cornell.edu, kinfai.mak@cornell.edu

**Controlling magnetism by purely electrical means is a key challenge to better information technology [1]. A number of different material systems, including ferromagnetic metals [2-4] and semiconductors [5], multiferroics [6-8] and magnetoelectric [9,10] materials, have been explored for electric-field control of magnetic properties. The recent discovery of two-dimensional van der Waals magnets [11,12] has opened a new door for electrical control of magnetism at the nanometer scale through the van der Waals heterostructure device platform [13]. Here we demonstrate control of magnetism in bilayer $CrI_3$, an antiferromagnetic semiconductor in its ground state [12], by application of small gate voltages using a field-effect device and by employing magnetic circular dichroism microscopy for detection. The applied electric field induces spin-dependent interlayer charge transfer, resulting in a large linear magnetoelectric effect, whose sign depends on the interlayer antiferromagnetic order. We also achieve reversible electrical switching of the interlayer magnetic order between the antiferromagnetic and ferromagnetic states in the vicinity of the interlayer spin-flip transition. The effect originates from the electric-field dependence of the interlayer exchange bias.**

      The recent discovery of two-dimensional (2D) van der Waals magnetic semiconductors, such as $CrI_3$ [12] and $Cr_2Ge_2Te_6$ [11], has attracted much attention. These materials provide unprecedented opportunities for studying magnetism in the 2D limit and engineering interface phenomena through van der Waals heterostructures [13,14]. In particular, $CrI_3$ is a model Ising ferromagnet with a strong out-of-plane anisotropy [15,16], whose magnetic properties remain robust down to the monolayer limit [12]. The low-temperature bulk structure of $CrI_3$ is rhombohedral [15,16], i.e. an ABCABC… stack of $CrI_3$ monolayers, in which the Cr atoms form a honeycomb structure in edge-sharing octahedral coordination by six I atoms (Fig. 1a). Below a Curie temperature of 68 K [17], the magnetic moment of $Cr^{3+}$ cations is aligned within each monolayer and between the layers in the out-of-plane direction by superexchange interactions through the I⁻ anions [15,16]. Intriguing layer-dependent magnetic order has been recently reported in atomically thin $CrI_3$ films [12] -- whereas monolayer and trilayer $CrI_3$ remain ferromagnetic (FM), bilayer $CrI_3$ becomes antiferromagnetic (AFM) with FM monolayers coupled antiferromagnetically (Fig. 1b). In contrast to FM monolayers and trilayers, the AFM

bilayers present a unique possibility for efficient electrical control of magnetism by the linear magneto-electric (ME) effect [18]. The linear ME effect, induction of the magnetization (polarization) by an electric (magnetic) field, requires breaking of both time-reversal and spatial-inversion symmetries [1,19-21]. The latter is satisfied only in AFM bilayers considering the magnetic symmetry [15,18,19] although spatial inversion is a fundamental crystal symmetry for $CrI_3$ of any thickness in the rhombohedral structure [15].

We fabricated dual-gate bilayer $CrI_3$ field-effect devices to investigate the electric-field effect on its magnetic order (details see Methods). In short, $CrI_3$ bilayers were exfoliated from bulk crystals and encapsulated between hexagonal boron nitride (hBN) substrates by the layer-by-layer dry transfer method [22,23]. Few-layer graphene was used as both top and bottom gate electrodes and contact electrodes. The device schematic and optical image of representative devices are shown in Supplementary Fig. S1. These devices with a total thickness of 40 nm allowed the application of giant out-of-plane electric fields ($E \sim 1$ V/nm) by moderate gate voltages ($\sim 30$ V). The dual-gate structure also allowed the independent control of the net doping density and electric field on bilayer $CrI_3$ [24]. Since no doping effect was observed within experimental uncertainty (Supplementary Sect. 3), we consider the field effect only. To probe the magnetic order, we employed the magnetic circular dichroism (MCD) microscopy with a HeNe laser at 633 nm (see Methods). The photon energy is near the absorption edge in $CrI_3$ [12] and the MCD signal is proportional to the sample's magnetization $M$.

Figure 1c illustrates the MCD signal of bilayer $CrI_3$ as a function of out-of-plane magnetic field $\mu_0 H$ under nearly zero electric field ($\mu_0$ is the vacuum permeability). At low temperatures, the observed dependence is consistent with the reported result [12]. Namely, for small magnetic fields no MCD signal was observed and the bilayer is in the AFM phase. A sharp rise in the MCD signal, corresponding to a spin-flip transition into the FM phase, was observed at a moderate field strength of $\mu_0 H_C \approx 0.4$ T because of the relatively weak interlayer exchange interaction. The occurrence of hysteresis upon forward and backward sweeps of the magnetic field suggests the first-order nature of the transition at 4 K, similar to the behavior in bulk $FeCl_2$ [25], which is an interlayer AFM with FM monolayers [15,25]. The result also suggests that in the AFM phase bilayer $CrI_3$ has two distinct spin configurations [12], which are time-reversal copies of one another and can be prepared by raising the magnetic field above $\mu_0 H_C$ (see below for direct experimental evidence). At higher temperatures, the spin-flip transition broadens and occurs at lower critical fields. The contour plot in Fig. 1d for the MCD signal as a function of magnetic field and temperature clearly shows the distinct FM, AFM and paramagnetic (PM) phases. The dashed line is the temperature dependence of the critical field for the spin-flip transition with the error bars denoting the transition width (detailed temperature dependence please see Supplementary Sect. 2). The critical field drops to zero around $T_C \approx 57$ K (vertical dotted line). We note that a small variation in the critical fields was observed in different devices, but their temperature dependence (Fig. 1d) and electric-field dependence (Fig. 2c) are practically identical for all devices.

Figure 2 shows the effect of an externally applied electric field on the magnetic properties of bilayer $CrI_3$. Three interesting features are observed in the magnetic-field dependence of the MCD signal in Fig. 2a. First in the AFM phase, the electric field $E$ induces a constant magnetization that increases with $E$ and has hysteresis in magnetic field sweeps. Magnetization as large as ~ 30% of the saturation magnetization $M_0$ was observed. Second, the spin-flip transition is pushed out to larger magnetic fields when $E$ is applied. Finally in the FM phase, $M_0$ is nearly independent of $E$. These observations are summarized in Fig. 2b and 2c. Figure 2c shows that the critical magnetic field (for both forward and backward sweeps of the magnetic field) increases with $E$ and reaches a maximum around 0.5 V/nm. Figure 2b illustrates the electric-field dependence of the change of magnetization from its value at zero electric field. It is expressed both as a relative change $\Delta M/M_0$ (left axis) and an absolute change $\Delta M$ in sheet magnetization (right axis) by assuming each $Cr^{3+}$ cation carry a magnetic moment of $\approx 3\mu_B$ ($\mu_B$ denoting the Bohr magneton) under saturation [15,16] (Methods). In the FM phase, $\Delta M/M_0$ (measured at 1 T) is negligible under small $E$'s and decreases nonlinearly with $E$. In contrast, in the AFM phase a substantially larger $\Delta M/M_0$ (measured at 0 T) is observed, which depends linearly on $E$ and changes sign at $E = 0$ V/nm (lines are linear fits). There are also two magnetization values of opposite sign for any nonzero $E$. They arise from the two AFM configurations as mentioned above. We can quantify this effect by using the linear ME coefficient $\alpha_{zz}$ [1,19-21], which relates $\Delta M$ to the applied vertical electric field as $\mu_0 \Delta M \equiv \alpha_{zz} E$. We obtained a sheet ME coefficient of $\alpha_{zz} \approx \pm 10^{-19}$ s for AFM bilayer CrI$_3$. The equivalent volumetric ME coefficient $\alpha_{zz}/2t \approx \pm 100$ ps/m ($t \approx 1$ nm denoting the interlayer separation in bilayer CrI$_3$ [15,16]) is comparable to the largest among known values for single-phase materials [26].

To compare the ME response of bilayer CrI$_3$ in different phases, we also computed the ratio $\mu_0 \Delta M/E$ using the experimental $M(H)$ results at 0.81 V/nm and 0 V/nm (Fig. 3). In the AFM phase, the ratio is just the linear ME coefficient, which does not depend on magnetic field. In the FM phase, the ME response is substantially smaller. A large enhancement of the response is observed near the critical magnetic field. This can be understood from the electric-field dependence of $H_C$ (Fig. 2c). Indeed, the application of an electric field can tip the balance between the AFM and FM phases and cause the spin-flip transition and a large change in the sample's magnetization. As we demonstrate below, such an enhanced ME response could be employed for electrical switching of magnetic order. In Fig. 3, we also include the result of a control experiment on monolayer CrI$_3$. The behavior of monolayer and bilayer CrI$_3$ is diametrically different. Negligible ME response was observed in monolayer CrI$_3$. (More data on monolayer CrI$_3$ and bilayer CrI$_3$ at different temperatures are included in Supplementary Sect. 8 and 4, respectively.)

The observed ME effect in bilayer CrI$_3$ can be understood considering the material's magnetic symmetry [19]. The time-reversal symmetry is broken in both the FM and AFM phases. In the FM phase, the spatial-inversion symmetry is present [15] so that no linear ME effect is allowed. And by the same token, the effect is not allowed in centrosymmetric monolayer CrI$_3$. We note that a nonlinear ME effect can still occur [20], which is indeed observed in Fig. 2b under large electric fields. In the AFM phase, however, the spatial-inversion symmetry is broken (Fig. 1b) and a nonzero linear ME tensor $\overleftrightarrow{\alpha}$ is consequently allowed [18]. In particular, the $\alpha_{zz}$ component is directly proportional to the AFM order parameter [9,27,28] $M_t - M_b$ ($M_t$ and $M_b$ denoting the sheet magnetization of the top and bottom layer, respectively). Since there are two distinct configurations of the AFM state with opposite AFM order parameters (inset, Fig. 1c), two $\alpha_{zz}$'s of opposite sign are expected. This is similar to the case of bulk DyPO$_4$ [28]. The observation of two $\alpha_{zz}$'s of opposite sign is thus an experimental verification of two distinct AFM configurations in bilayer CrI$_3$. As discussed above, these two AFM configurations can be prepared by raising the magnetic field above $\mu_0 H_C$ (Fig. 1c). They can also be prepared by cooling the sample from above $T_C$ under both magnetic and electric fields, known as ME annealing [9] (Supplementary Sect. 5). The microscopic mechanism for the ME effect in bilayer CrI$_3$, however, remains unknown. A plausible mechanism involves charge transfer [18], for instance, from the top layer to the bottom layer under an up electric field $E$ (Fig. 1b). Because of the intralayer FM exchange coupling, the transferred particles would align their spins parallel to the existing ones to minimize the system's total free energy. The AFM bilayer thus acquires an up net magnetization, corresponding to a positive $\alpha_{zz}$. For the second AFM configuration with a reversed AFM order parameter, the net magnetization would point down, corresponding to a negative $\alpha_{zz}$. This is fully consistent with experiment. We can also estimate the magnitude of the ME coefficient based on this picture by assuming each transferred electron carry a magnetic moment $\mu_B$. Under an applied field of 0.81 V/nm, the net carrier density in bilayer CrI$_3$ is estimated from the parallel plate capacitance model to be $\sim 10^{14}$ cm$^{-2}$ (see Supplementary Sect. 7 for details). This gives rise to a volumetric ME coefficient of $\sim 10$ ps/m, which is compatible with the experimental result.

The observed electric-field dependence of the critical magnetic field (Fig. 2c) for the spin-flip transition from the AFM to the FM phase is more complex. Several effects could contribute to it. The electric field can lower the free energy of the AFM phase through the ME effect [9, 20] and lead to a higher critical field for the spin-flip transition. The electric field can also change the interlayer exchange coupling and the free energies through the electron wave function overlap in the vertical direction. Furthermore, the electric field can change the magnetic anisotropy, for instance, through the Rashba spin-orbit interaction [18] or through changing the electron occupancy in the 3d orbitals [3, 4]. As discussed in ref. [25], magnetic anisotropy can act as an energy barrier for the spin-flip transition at $H_C$, leading to the occurrence of hysteresis. Below we evaluate the importance of the ME effect in the spin-flip transition. First at zero electric field, the free energy per unit area at low temperatures can be expressed as [9] $F = 2F_0 - J$ in the AFM phase, and $F = 2F_0 + J - \mu_0 M_0 (H - M_0/2t)$ in the FM phase ($M_0 > 0$). (Higher order terms in $H$ have been ignored.) Here $F_0$ denotes the free energy of the constituent monolayer, which is identical in the FM and AFM phases; $J$ ($> 0$) is the interlayer exchange energy, which adds to the free energy $+J$ in the FM phase since the spins are parallel, and $-J$ in the AFM phase since the spins are anti-parallel; the magnetic energy under an applied vertical magnetic field $\mu_0 H$ is nonzero only in the FM phase. The critical field is thus determined as $\mu_0 H_C = \frac{2J + \mu_0 M_0^2/2t}{M_0}$ by setting the two free energies equal. The expression is reminiscent of that for the exchange bias field at the FM-AFM interfaces [9, 29]. In fact, the critical field here can be regarded as the exchange bias field provided by one of the FM monolayers. From the measured critical magnetic field value we estimated the exchange coupling energy to be $J \approx 40$ μJm$^{-2}$. Next when an electric field $E$ is turned on, the AFM phase acquires an additional ME energy $\approx -\alpha_{zz} E H$ [9, 20] (the $E^2$ term would cancel), which leads to a new critical magnetic field

$$\mu_0 H_C = \frac{2J + \mu_0 M_0^2/2t}{M_0 - \alpha_{zz} E/\mu_0}. \tag{1}$$

The prediction of Eq. (1) using the measured values for the parameters is shown in Fig. 2c as a blue line. The simple model captures the correct magnitude for the electric field-dependent $H_C$. However, it predicts a monotonic increase of $H_C$ with increasing $E$ and cannot explain the observed decrease of $H_C$ for $E > 0.5$ V/nm. We thus conclude that the ME effect plays a major role in determining the critical field for the spin-flip transition, but other effects such as the electric field-induced change in the interlayer coupling energy and/or magnetic anisotropy need to be considered to fully explain the observation in future studies.

Finally, we demonstrate pure electrical switching of magnetic order in bilayer $CrI_3$ by taking advantage of the large ME response near the critical field. Figure 4a shows the sheet magnetization $M$ (right axis) and the normalized magnetization $M/M_0$ (left axis) of bilayer $CrI_3$ as a function of forward and backward sweeps of the electric field under two fixed magnetic fields ($\pm 0.44$ T, for results under more magnetic fields see Supplementary Sect. 6). Remarkably, the electric field switches the material from a FM phase (< 0.2 V/nm) to an AFM phase (> 0.7 V/nm). And the magnetization varies from ~ 0.8 $M_0$ to ~ 0.2 $M_0$. The hysteresis, again, indicates the first-order nature of the transition [25]. In Fig. 4b, we further show that the switching operation can be repeated many times by turning the electric field on/off periodically. The magnetization is seen to follow the applied electric field with no sign of fatigue. Our results demonstrate the unique potential of 2D van der Waals magnets for electrically controlled nonvolatile memories and spintronic and valleytronic devices through proximity coupling in van der Waals heterostructures [13,14].

**Methods**

**Device fabrication and characterization.** Dual-gate field-effect devices of atomically thin $CrI_3$ with hexagonal boron nitride (hBN) as gate dielectric and few-layer graphene as gate and contact electrodes were fabricated by the lay-by-layer dry transfer method [22,23]. Atomically thin flakes of $CrI_3$ (HQ Graphene), hBN and graphene were mechanically exfoliated from their bulk crystals onto silicon substrates covered by a 300 nm thermal oxide layer. Due to the instability of $CrI_3$ in air, $CrI_3$ was handled only inside a glovebox under controlled atmosphere with less than one part per million oxygen and moisture [11,12]. In ambient atmosphere, stamps for transfer, consisting of a thin layer of polycarbonate (PC) on polydimethylsiloxane (PDMS) supported by a glass slide, were prepared. The bottom graphene gate electrode and hBN gate dielectric were picked up by a stamp and released onto a silicon substrate with pre-patterned gold electrodes. The residual PC was dissolved in chloroform. The top graphene gate electrode, hBN gate dielectric and graphene contact electrodes were picked up by another stamp and introduced into the glovebox together with the substrate with the bottom gate. Inside the glovebox, the stamp picked up $CrI_3$ and released the entire stack onto the substrate with the bottom gate. After this step, the device was safe to be removed from the glovebox since $CrI_3$ was encapsulated by hBN. The residual PC on the device surface was dissolved in chloroform before optical measurements.

The thickness of atomically thin materials was initially estimated from their optical reflectance contrast on silicon substrates and later verified by the atomic force microscopy (AFM) measurements. The layer thickness of CrI$_3$ was further confirmed from the magnetization measurement under a varying out-of-plane magnetic field. The typical thickness of hBN gate dielectric was ~ 20 nm and nearly identical for the top and bottom gates. The applied electric field was varied by changing the difference between the top and bottom gate voltages, referred to simply as the gate voltage in the main text. More details on the device structure are provided in Supplementary Sect. 1 and 7. Our devices all showed a built-in electric field, likely due to the asymmetry in the fabrication procedure for the top and bottom gates. An electric field at a level of ~ 0.4 V/nm was typically required to cancel the built-in field. We subtracted this value from the applied electric field in all presented results so that in the pristine state bilayer CrI$_3$ is an antiferromagnet (i.e. zero net magnetization at zero magnetic field).

**Magnetic circular dichroism (MCD) microscopy.** The MCD measurements were performed in an Attocube closed-cycle cryostat (attoDry1000) down to 4 K and up to 1 Tesla in the out-of-plane direction with a HeNe laser at 633 nm. Optical radiation with power ~ 5 μW was coupled into and out of the system using free-space optics. A high numerical aperture (NA = 0.8) objective was used to focus the excitation beam onto the device with a sub-micron spot size. The optical excitation was modulated between left and right circular polarization by a photoelastic modulator (PEM) at 50.1 kHz. The reflected beam was collected by the same objective and detected by a photodiode. The MCD was determined as the ratio of the ac component at 50.1 kHz (measured by a lock-in amplifier) and the dc component (measured by a multimeter) of the reflected light intensity.

**Data analysis.** The critical magnetic field for the spin-flip transition was determined from the peak position of the differential magnetic susceptibility [30], which was calculated numerically from the measured magnetic-field dependence of the MCD signal. The full-width-at-half-maximum (FWHM) of the peak was taken to be the transition width as shown by the error bars in Fig. 1d and 2c. The MCD signal was converted to sheet magnetization by assuming that the MCD signal is linearly proportional to sheet magnetization and the saturation magnetization is $M_0$ = 0.274 mA. The latter was obtained by assuming that under saturation each Cr$^{3+}$ cation carries a magnetic moment of $3\mu_B$ [15,16]. The density of Cr was calculated using the crystallographic data of bulk CrI$_3$ (space group $R\bar{3}$ with unit cell parameters of $a = 6.867$ Å, $b = 6.867$ Å, $c = 19.807$ Å and $\beta = 90°$) [15,16].

Figures

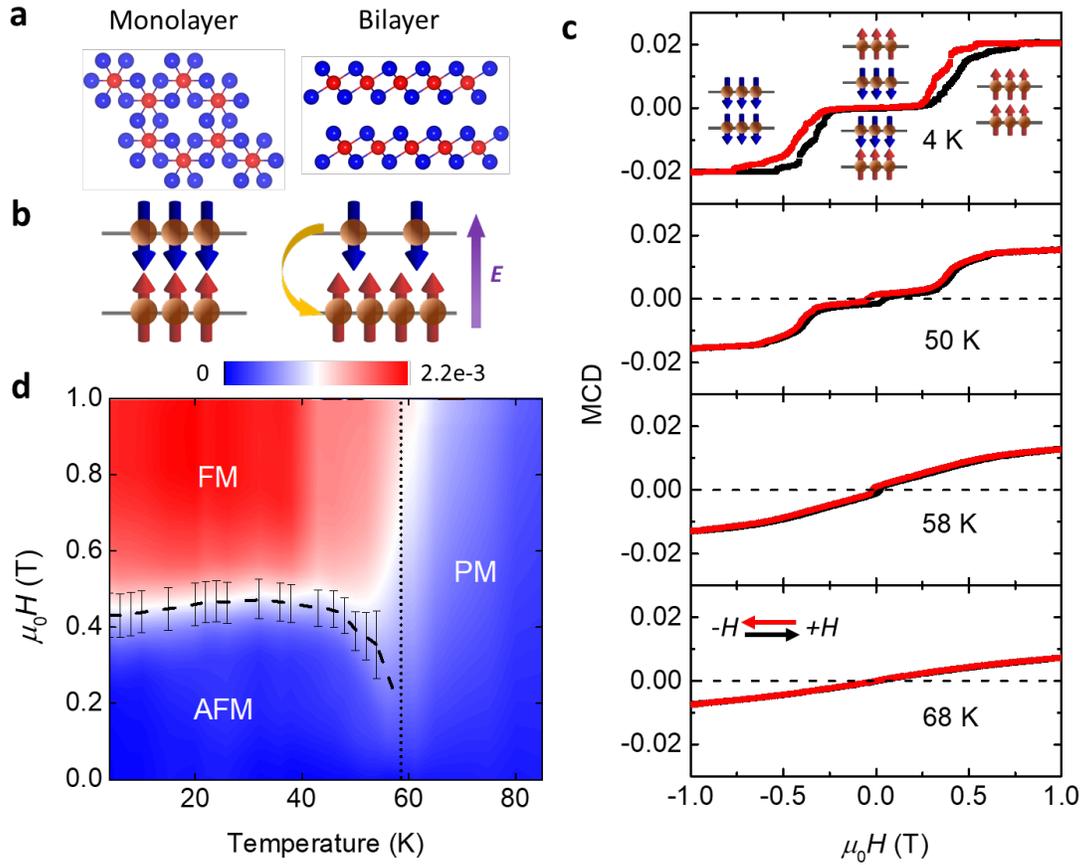

**Figure 1 | Crystal structure and magnetic phase diagram of bilayer CrI$_3$. a,** Top view of monolayer CrI$_3$, where Cr atoms (red balls) form a honeycomb structure in edge-sharing octahedral coordination by six I atoms (blue balls) and side view of bilayer CrI$_3$ of the rhombohedral stacking order. **b,** AFM bilayer CrI$_3$ consists of two FM monolayers with antiferromagnetic interlayer coupling. The net magnetization is zero. Spin-dependent charge transfer between the top and bottom layer under a vertical electric field $E$ leads to a nonzero net magnetization. **c,** MCD signal as a function of applied magnetic field at different temperatures. Black and red lines are for the forward and backward sweeps of the magnetic field. Insets depict the magnetic ground states of bilayer CrI$_3$ under different magnetic fields. **d,** $H$-$T$ phase diagram of the magnetic order in bilayer CrI$_3$ determined from the magnitude of the MCD. FM, AFM and PM denote the ferromagnetic, antiferromagnetic and paramagnetic phase, respectively. The dashed line with error bars is the temperature dependence of the critical magnetic field of the spin-flip transition and the transition width. The dotted line indicates the critical temperature for the spin-flip transition.

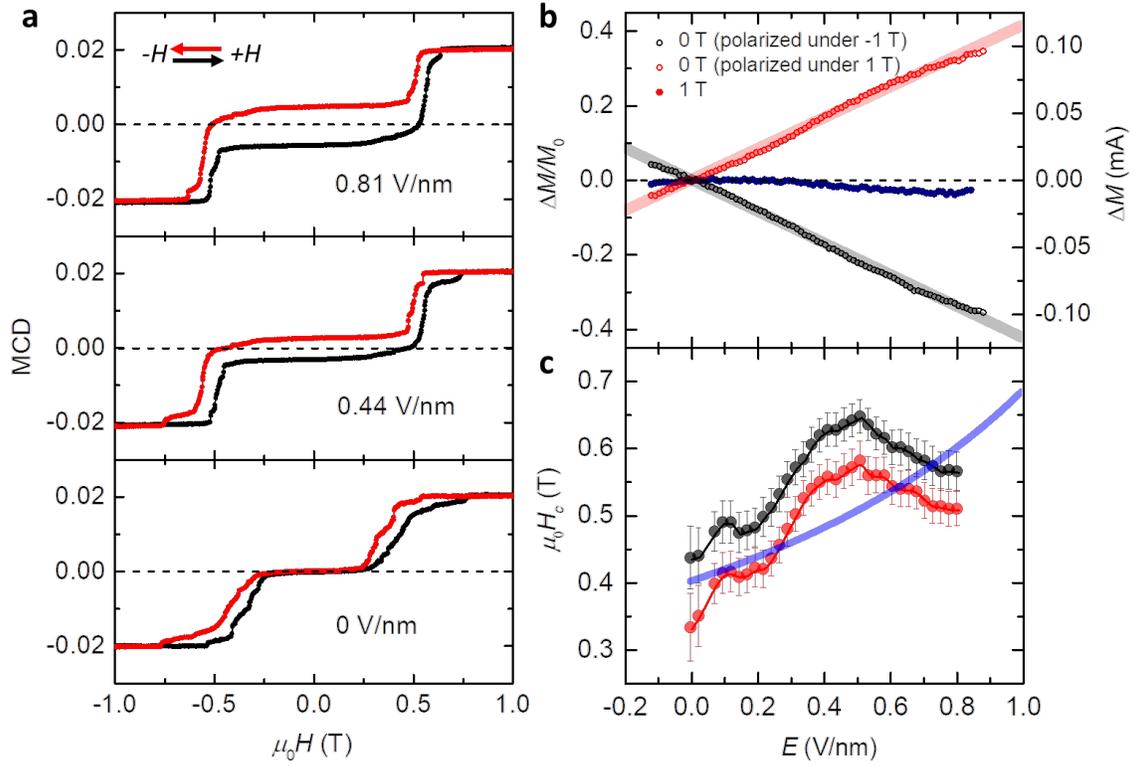

**Figure 2 | Linear magnetoelectric effect in AFM bilayer $CrI_3$. a,** MCD signal as a function of magnetic field under representative electric fields at 4 K. **b,** Relative and absolute change in the sheet magnetization ($\Delta M/M_0$ and $\Delta M$) as a function of applied electric field measured under a fixed magnetic field at 0 T (open symbols) and 1 T (filled symbols). The lines are linear fits to the data at 0 T. **c,** Critical magnetic field for the spin-flip transition (symbols) and the transition width (error bars) as a function of applied electric field. The line is the prediction of Eq. (1).

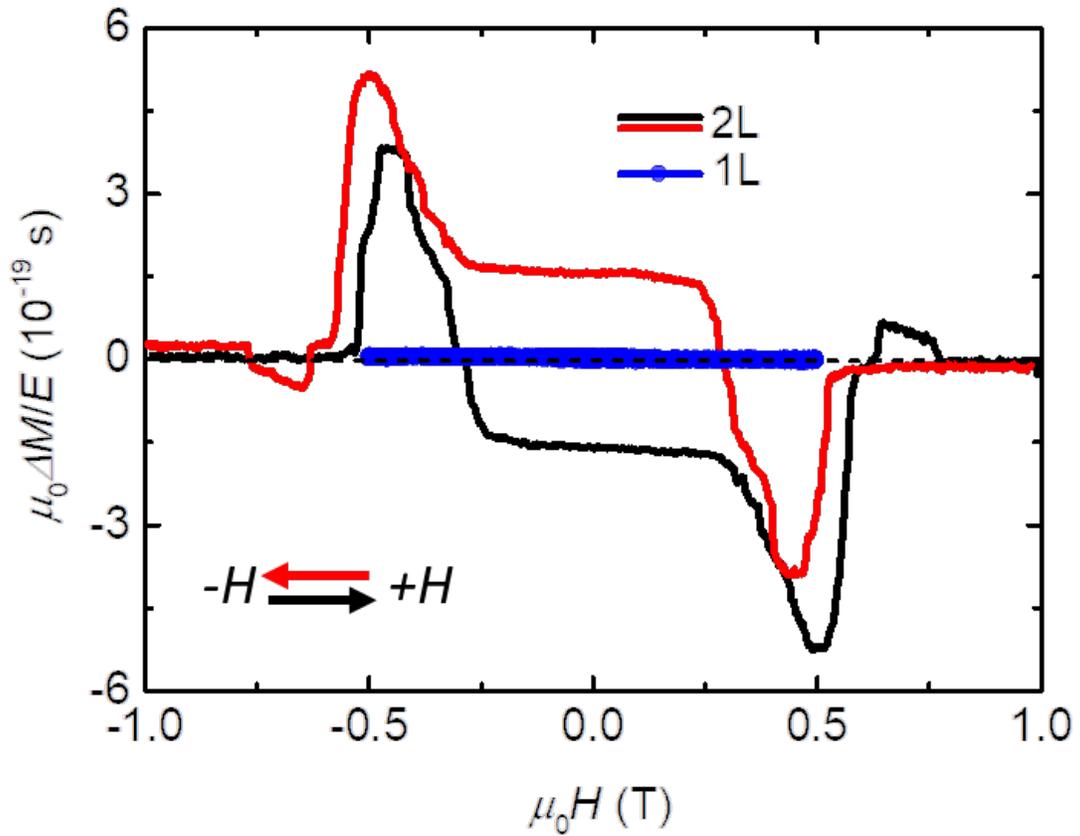

**Figure 3 | Magnetoelectric response of bilayer and monolayer CrI$_3$.** Rate of the induced change in the sheet magnetization by the applied electric field at 4 K obtained by subtracting *M-H* curves under 0.8 V/nm and 0 V/nm for bilayer CrI$_3$. Black and red solid lines are for forward and backward sweeps of the magnetic field. For monolayer CrI$_3$ (symbols), the *M-H* curves under 0.34 V/nm and - 0.34 V/nm were used in the calculation.

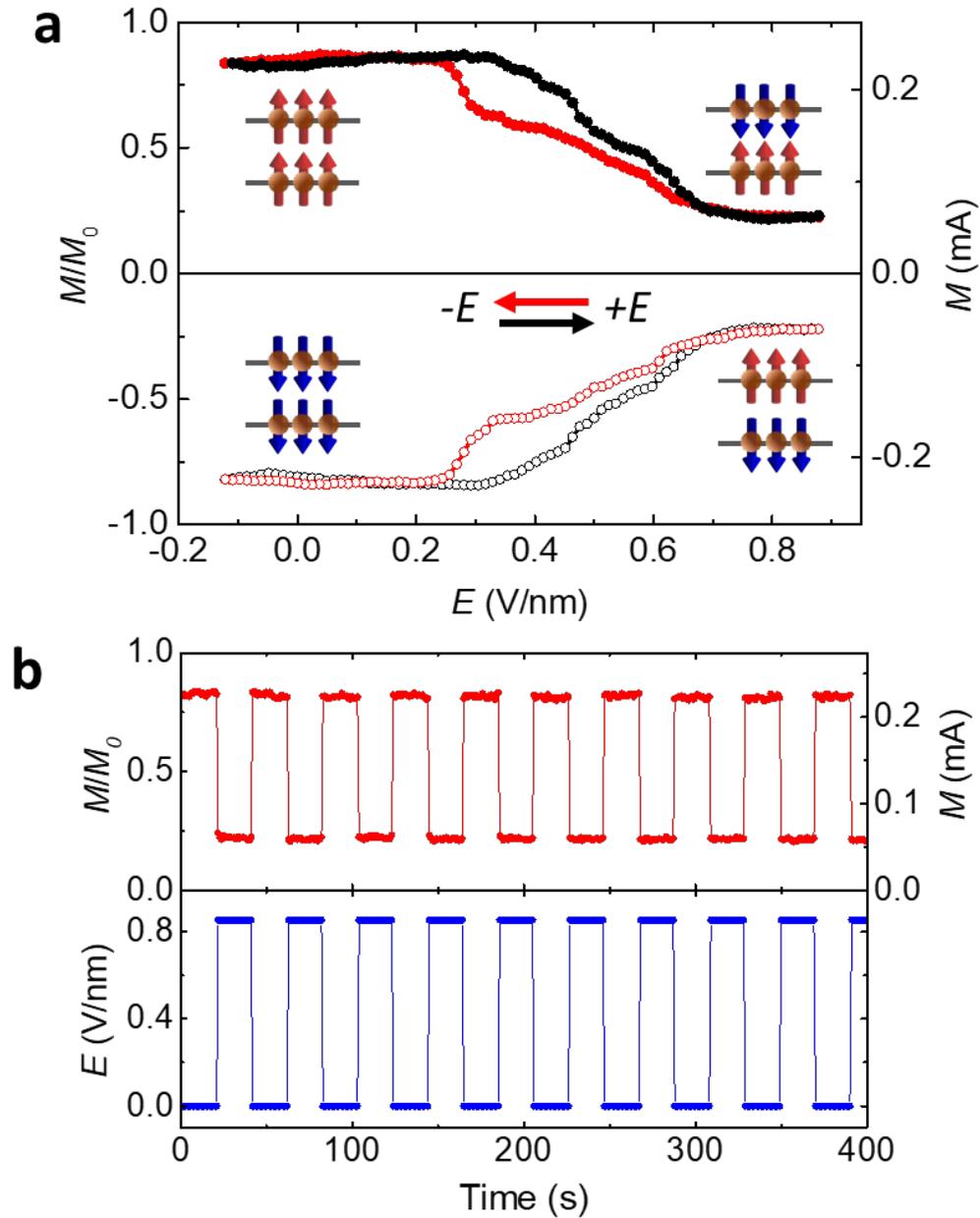

**Figure 4 | Electrical switching of magnetic order in bilayer CrI$_3$. a,** Magnetization $M$ (right) and normalized magnetization $M/M_0$ (left) as a function of applied electric field $E$ under fixed vertical magnetic fields near the critical value (filled symbols for 0.44 T and empty symbols for - 0.44 T, respectively). Black and red symbols are for forward and backward sweeps of $E$. Insets depict the magnetic states under different magnetic and electric fields. **b,** Repeated switching of the magnetic order (and magnetization) by the periodic application of an electric field under a constant magnetic field (0.44 T).